\documentclass[aps,axodraw,twocolumn,prl]{revtex4}
\usepackage{fancyhdr}
\usepackage{fancybox}
\usepackage{amsmath,amssymb,graphicx}

\newcommand{\beq}{\begin{eqnarray}}
\newcommand{\eeq}{\end{eqnarray}}

\newcommand{\drawsquare}[2]{\hbox{%
\rule{#2pt}{#1pt}\hskip-#2pt
\rule{#1pt}{#2pt}\hskip-#1pt
\rule[#1pt]{#1pt}{#2pt}}\rule[#1pt]{#2pt}{#2pt}\hskip-#2pt
\rule{#2pt}{#1pt}}

\newcommand{\fund}{\drawsquare{6.5}{0.4}}

\newcommand{\asymm}{\raisebox{-3.5pt}{\drawsquare{6.5}{0.4}\hskip-6.9pt%
        \raisebox{6.5pt}{\drawsquare{6.5}{0.4}}}}

\newcommand{\ov}{\overline}

\begin{document}


\title{ A Spin-1 Top Quark Superpartner} 


\author{ Haiying Cai, Hsin-Chia Cheng, and John Terning}


\address{ Department of Physics, University of California, Davis,
CA
95616.} 


\begin{abstract}
We construct a supersymmetric model
where the left-handed top and bottom quarks are mainly the gauginos of a vector
supermultiplet and hence their superpartners are spin-1. The right-handed 
top quark is unified with the Higgs, the top Yukawa arises from the gaugino
coupling.

\end{abstract}

\maketitle


In the Standard Model (SM), the Higgs mass-squared receives
quadratically divergent corrections from interactions with other SM fields.
These contributions should be cut off at scales not much higher
than the electroweak symmetry breaking scale so that the electroweak scale is stable.
The largest contribution comes from a top quark loop through the top Yukawa coupling.
Requiring that there is no more than 10\% fine tuning in the Higgs mass parameter,
the top loop should be cut off below $\sim$ 2 TeV. This implies new physics at
or below the TeV scale, which will be explored at the Large Hadron Collider (LHC).

The most popular and promising candidate for new physics at the
TeV scale is supersymmetry (SUSY). In the Minimal Supersymmetric
Standard Model (MSSM), there is a superpartner for each SM particle
with the opposite spin-statistics, and the quadratic radiative
corrections are cancelled between bosons and fermions.
If SUSY is responsible for stabilizing the electroweak scale, the SUSY-breaking masses 
of the superpartners are expected to be around the TeV scale. In particular,
the superpartners of the top quark, which are scalar particles in the MSSM,
are required to have TeV or smaller masses in order to avoid fine-tuning.
The top's superpartner, the top squark, is necessarily colored and
can be copiously produced at the LHC.

In recent years, several other possibilities of cancelling the
quadratic divergences to the Higgs mass-squared were discovered.
In little Higgs theories~\cite{LH}, the one-loop quadratic
divergences from the SM particles are cancelled by new particles at
the TeV scale with the {\it same\/} spins as the corresponding SM
particles. The top partner which cuts off the top loop is a
spin-1/2 fermion in this case. Again it also carries color so  it
will be copiously produced at the LHC if it exists. It is also
possible to cancel the top loop by new particles which are not
charged under color, but another $SU(3)$ gauge group, as in the twin Higgs model~\cite{twin}
and folded supersymmetry~\cite{folded}. There can be very exotic
phenomenology associated with these models.

 In  this Letter we will explore the possibility that the top partner responsible
for cancelling the top loop is a spin-1 particle. Such a possibility should
not be unexpected, as SUSY relates particles with spins that differ by 1/2.
Usually in supersymmetric models both the left- and right-handed top quarks belong to
chiral supermultiplets and their superpartners are spin-0 particles.
One can consider the alternative possibility of assigning the top to a vector
supermultiplet so that its superpartner is a spin-1 vector boson. In this letter
we construct a relatively simple model where the superpartner of the left-handed
top quark is a spin-1 vector particle. (The superpartner of the right-handed
top quark is still a scalar.) There are many interesting aspects of this model
and we will briefly discuss its phenomenology.


One immediate challenge to assign the top to a vector supermultiplet
is that the vector supermultiplet transforms as an adjoint representation
under some gauge group while the top quarks are fundamentals under the SM gauge group.
However, if we consider an enlarged gauge group such as $SU(5)$,
it contains gauge bosons transforming as ({\bf 3}, {\bf 2}) under the $SU(3)\times SU(2)$ subgroup,
and we can try to identify them as the superpartner of the left-handed top quark.
By identifying the left-handed top quark as a gaugino, the top Yukawa
coupling should come from a gaugino interaction, which implies that the right-handed
top and the Higgs should be unified into a chiral supermultiplet transforming
under the $SU(5)$ gauge group.

Our model is based on the gauge group $SU(3) \times SU(2) \times U(1)_H \times U(1)_V \times SU(5)$.
It is broken at the TeV scale down to the diagonal SM gauge group $SU(3)_C\times SU(2)_L \times U(1)_Y$
by the nonzero vacuum expectation values (VEVs) of the fields $\Phi_3$, $\ov{\Phi}_3$, 
$\Phi_2$, $\ov{\Phi}_2$.
We include three generations of matter fields (quarks and leptons)
which transform under $SU(3)\times SU(2) \times U(1)_H$.
The two Higgs doublets belong to ${\bf 5}$ and $\bar{\bf 5}$ of the $SU(5)$ gauge group.
They also carry $U(1)$ charges in order to arrange the correct hypercharges after
symmetry breaking. The field content is given in Table~\ref{tab:content}.
\begin{table}
\begin{center}
\begin{tabular}{c|ccr|rc|c}
 & $SU(3)$ & $SU(2)$  & $U(1)_H$ & $U(1)_V$ & $SU(5)$ & $H+V+a T_{24}$\\
\hline $Q_i$ &\fund & \fund & $\frac{1}{6}$  & $0$\hphantom{0}   & ${\bf 1}$&$
\frac{1}{6}$ $\vphantom{\raisebox{3pt}{\asymm}}$
\\
$\overline{u}_i$ & $\overline{\fund}$     & ${\bf 1}$
 & $-\frac{2}{3}$& $0$\hphantom{0}   &${\bf 1}$ &$-\frac{2}{3}$  $\vphantom{\raisebox{3pt}{\asymm}}$
\\
$\overline{d}_i$ & $\overline{\fund}$     & ${\bf 1}$
 & $\frac{1}{3}$& $0$\hphantom{0}   &${\bf 1}$ & $\frac{1}{3}$$\vphantom{\raisebox{3pt}{\asymm}}$
\\
$L_i$ & ${\bf 1}$ & \fund & $-\frac{1}{2}$& $0$\hphantom{0}   &${\bf 1}$
&$-\frac{1}{2}$ $\vphantom{\raisebox{3pt}{\asymm}}$
\\
$\overline{e}_i$ &${\bf 1}$     & ${\bf 1}$
 & $1$& $0$\hphantom{0}   &${\bf 1}$ &$1$  $\vphantom{\raisebox{3pt}{\asymm}}$
\\
$H$ & ${\bf 1}$ & ${\bf 1}$ & $\frac{1}{2}$ & $\frac{1}{10}$\hphantom{0}   &\fund
&$(\frac{2}{3},\frac{1}{2})$
$\vphantom{\raisebox{3pt}{\asymm}}$
\\
${\overline H}$ & ${\bf 1}$ &  ${\bf 1}$ &  $-\frac{1}{2}$  & $-\frac{1}{10}$\hphantom{0}
&${\overline \fund}$ &$(-\frac{2}{3},-\frac{1}{2})$
$\vphantom{\raisebox{3pt}{\asymm}}$\\

$\Phi_3$ & ${\overline \fund} $& ${\bf 1}$  & $-\frac{1}{6}$ & $\frac{1}{10}$\hphantom{0}    &
\fund  &$(0,-\frac{1}{6})$
$\vphantom{\raisebox{3pt}{\asymm}}$\\
$\Phi_2$ & ${\bf 1}$ & ${\overline \fund}$ & $0$ & $\frac{1}{10}$\hphantom{0} 
&\fund &$(\frac{1}{6},0)$
$\vphantom{\raisebox{3pt}{\asymm}}$\\

${\overline \Phi_3}$ &  \fund   & ${\bf 1}$ & $\frac{1}{6}$ & -$\frac{1}{10}$\hphantom{0} 
&${\overline \fund}$ &$(0,\frac{1}{6})$
$\vphantom{\raisebox{3pt}{\asymm}}$\\

${\overline \Phi_2}$ & ${\bf 1}$ & \fund & $0$  & $-\frac{1}{10}$\hphantom{0}
&${\overline \fund}$&$(-\frac{1}{6},0)$
$\vphantom{\raisebox{3pt}{\asymm}}$\\

\end{tabular}
\end{center}
\caption{\label{tab:content} The field content of the model.
The last column is the hypercharge of the diagonal hypercharge $U(1)_Y$,
where $Y=H+V+aT_{24} $, with $a= 1/\sqrt{15}$.}
\end{table}

The triplet partners of the Higgs doublets in $\ov{H}$ and $H$
have the same quantum numbers as the right-handed top and its conjugate.
We can write them as
\beq
{\overline H}=(\ov{T}, H_1), \quad
H=(\overline{T}^c, H_2). \label{H}
\eeq

The superpotential is given by
\beq
\label{eq:superpotential}
 W&=&
y_1 Q_3 \Phi_3 {\overline \Phi_2}+\mu_3 \Phi_3 {\overline
\Phi_3}+\mu_2  \Phi_2 {\overline \Phi_2}
\\
& +& y_2 \ov{u}_3 H {\overline \Phi_3}
+\mu_H H {\overline
H}+{Y}_{Uij} \, Q_i\, {\overline u}_j\,   {\overline \Phi_2} H 
 \nonumber
\\
&+& Y_{Dij}\,Q_i
\, {\overline d}_j  \Phi_2 {\overline H}+Y_{Eij} \,L_i \, {\overline e}_j
\Phi_2 {\overline H}.\nonumber
\eeq
There are also the usual soft SUSY-breaking terms, including the
gaugino masses, scalar masses, $A$-terms and $B$-terms. We do
not specify the origin of these soft SUSY-breaking terms as it is
not essential for the discussion here.  We assume that due to SUSY breaking the potential for
$\Phi_j$, $\ov{\Phi}_j$ is unstable at the origin so that they get the following nonzero VEVs,
breaking $SU(3) \times SU(2)\times U(1)_H$ and $SU(5)\times U(1)_V$ down to the
diagonal SM gauge group,
\beq 
\langle \Phi_3 \rangle\hspace{-3pt}  &=&\hspace{-3pt}  \left( \begin{array}{ccccc}
f_3\hspace{-3pt} &  0&0 &0 &0 \\
0 & f_3\hspace{-3pt} & 0&0 &0\\
0&0 & f_3\hspace{-3pt}&0 &0
\end{array} \right),\,
\langle \overline \Phi_3
\rangle^T\hspace{-2pt} = \hspace{-3pt} \left( \begin{array}{ccccc}
{\overline f_3}\hspace{-3pt} &  0&0 &0 &0 \\
0 &{\overline f_3}\hspace{-3pt} & 0&0 &0\\
0&0 &{\overline f_3}\hspace{-3pt} &0 &0
\end{array} \right),
\nonumber \\
\langle \Phi_2 \rangle\hspace{-3pt}  &=&\hspace{-3pt}  \left( \begin{array}{ccccc}
0 &0 &0 &f_2 \hspace{-3pt}&0 \\
0 &0 &0 &0 &f_2\hspace{-3pt}
\end{array} \right), 
\, \langle{\overline  \Phi_2}
\rangle^T\hspace{-2pt} =\hspace{-3pt}  \left( \begin{array}{ccccc}
0 &0 &0 &{\overline f_2}\hspace{-3pt} &0 \\
0 &0 &0 &0 &{\overline f_2}\hspace{-3pt}
\end{array} \right)\hspace{-1pt} . 
\label{PHIbar}
\eeq

The alignment of the VEVs of $\Phi_3$ and $\ov{\Phi}_2$ is ensured
by the $y_1 Q_3 \Phi_3 {\overline \Phi_2}$ term in the superpotential.
The VEVs $f_3$, $\bar{f}_3$, $f_2$, and $\bar{f}_2$ are not
equal in general as they depend on the soft SUSY breaking terms
and $\mu_2$, $\mu_3$. 
The unequal VEVs can generate a $U(1)_H$ $D$-term,
$(\hat{g}_{1H}/2)(\bar{f}_3^2 - f_3^2)$, and a 
$U(1)_V$ $D$-term,
$(\hat{g}_{1V}/10)(3(f_3^2 - \bar{f}_3^2) +2( f_2^2 - \bar{f}_2^2))$.
which in turn gives additional contributions to the scalar
masses of fields charged under the $U(1)$s. We assume that these
are small enough not to affect the spectrum dramatically.

With the above VEVs, the $\Phi$ fields split into
the following representations under
$SU(3)_C\times SU(2)_L \times U(1)_Y$:
\beq
\Phi_3&\rightarrow&(1,1,0)+({\bf 8},1,0)+({\bar {\bf 3}}, {\bf 2},-1/6)\nonumber\\
{\ov \Phi}_3&\rightarrow&(1,1,0)+({\bf 8},1,0)+({\bf 3},{\bf 2},1/6)\nonumber \\
\Phi_2&\rightarrow&({\bf 3},{\bf 2},1/6)+(1,1,0)+(1,{\bf 3},0)\nonumber\\
{\ov \Phi}_2&\rightarrow&({\bar{\bf 3}},{\bf 2},-1/6)+(1,1,0)+(1,{\bf 3},0)
\eeq

The gauge couplings for the SM $SU(3)_C\times SU(2)_L \times U(1)_Y$
gauge group are given by
\beq \frac{1}{{g_{2,3}^2 }} = \frac{1}{{\hat g_{2,3}^2 }} +
\frac{1}{{\hat g_5^2 }}, \quad\quad  \frac{1}{{g_{1}^2 }} =
\frac{1}{{\hat g_{1H}^2 }} + \frac{1}{{\hat g_{1V}^2 }} + \frac{1}{15{\hat g_5^2 }},
\eeq 
where $\hat g_i$ and $\hat g_5$ are the gauge couplings of the original
$SU(3)$, $SU(2)$, $U(1)_H$, $U(1)_V$ and $SU(5)$ gauge groups
respectively. 
There are two broken $U(1)$ gauge bosons whose masses are fixed by
\beq
{\mathcal L}\supset&& \frac{1}{2} \bigg\{ 6(f_3^2+\bar{f}_3^2)(\frac{\hat g_{1H} }{6}B_{1H}-\frac{\hat g_{1V}}{10}B_{1V}-\frac{\hat g_5}{\sqrt{15}}B_{24})^2\nonumber\\
&&+4(f_2^2+\bar{f}_2^2)(\frac{\hat g_{1V}}{10}B_{1V}-\frac{\sqrt{15}}{10}\hat g_5 B_{24})^2\bigg\}~.
\eeq
The masses of the other heavy gauge bosons after symmetry
breaking are given by,
\beq
 m_{G'}^2  &=& (\hat g_3^2  + \hat g_5^2 )(f_3^2  + \bar f_3^2 ) , \label{eq:gprime}\\
 m_{W'}^2  &=& (\hat g_2^2  + \hat g_5^2 )(f_2^2  + \bar f_2^2 ) , \label{eq:Wprime}\\
 m_{\vec Q}^2  &=& \frac{1}{2}\hat g_5^2 (f_3^2  + \bar f_3^2  + f_2^2  + \bar
 f_2^2).
\label{eq:vtopmass}
\eeq
The ${\vec Q}$ gauge boson will be the dominant superpartner of the
left-handed top  and bottom quark doublet $Q$ as we will see
later. We denote it with an arrow above the particle name 
to emphasize that it is a spin-1 superpartner as opposed to the usual tilde used for superpartners. 
In $SU(5)$ grand unified models these gauge bosons are
often denoted by $X$ and $Y$. For other particles beyond the MSSM, we use the notation 
that the particles with odd $R$-parity in a supermultiplet have a tilde above the particle name, and the particles with even
$R$-parity do not.

The Yukawa couplings for the light SM fermions arise
from the last three terms of the superpotential~(\ref{eq:superpotential}).
After substituting in the VEVs of the $\Phi_2$ and $\ov{\Phi}_2$ fields,
they become the usual Yukawa terms. The fact that they come from
nonrenormalizable interactions can explain why they are small.
Note that the scalar components of the $\Phi_2$ and  ${\ov \Phi_2}$
fields that get VEVs must have the same $R$-parity assignments
as the Higgs fields, i.e. $+$.
We will come back to the issue of the $R$-parity assignments later.

For the top quark, $Q_3$ and $\ov{u}_3$ mix with other
states of the same quantum numbers under the SM
gauge symmetry through the $y_1$, $y_2$, and $\mu$ terms.
The mass matrix for the fermions in the ({\bf 3}, {\bf 2}, 1/6) sector  is
\beq \begin{array}{c|ccccc} &\lambda & \Phi_{2t} &
 {\overline  \Phi}_{3t} & Q_3 \\ \hline
\bar \lambda & M_5 &\hat g_5 f_2 &\hat g_5 \bar f_3 &0 \\
\Phi_{3t} &\hat g_5 f_3 &0 &\mu_3 &y_1 \bar f_2 \\
{\overline  \Phi}_{2t} &\hat g_5 \bar f_2 &\mu_2 &0 &y_1 f_3
\end{array},\eeq
where $\lambda$,  and $\bar \lambda$ are the gauginos
corresponding to the broken generators in $SU(5)$ which are in the
vector supermultiplets with ${\vec Q}$ and  ${\vec {
Q}}^*$; $M_5$ is the SUSY-breaking gaugino mass for all $SU(5)$
gauginos;  $\Phi_{2t}$, ${\overline  \Phi}_{3t}$ are the color
triplet fermions from $\Phi_2$, $\ov{\Phi}_3$, and $\Phi_{3t}$, ${\overline
\Phi}_{2t}$ are the color anti-triplet fermions. The massless left-handed top quark is
a linear combination of $\lambda$, $\Phi_{2t}$, ${\overline
\Phi}_{3t}$, and $Q_3$. For $M_5 \ll \hat g_5
f_2$, $\hat g_5 f_3 \ll \mu_3\, (\hat g_5 \bar f_2)$, and $\hat
g_5 \bar f_2 \ll \mu_2$, the massless left-handed top quark state
is mostly made of the gaugino $\lambda$. For example, if we take
$\bar f_2 = 1.5$~TeV, $f_2=1.7$~TeV, $\bar f_3 =0.6$~TeV, $f_3 =0.4$~TeV, $M_5 =0.7$~TeV, 
$\mu_2=5$~TeV, $\mu_3= 2$~TeV, $\hat g_5 =1.2$, and $y_1=1.5$, then the
massless combination is
\beq 
\label{eq:lefttop}
Q\equiv(t,\, b)_L \approx 0.93 \lambda -0.31 \Phi_{2t}
- 0.02 {\overline \Phi}_{3t} - 0.18 Q_3. \eeq

For the right-handed top quark, $\ov{u}_3$ mixes with the colored
fermion component of $\ov{H}$ through the $y_2$ term in
the superpotential Eq.(\ref{eq:superpotential}).
The mass matrix is:
\begin{center}
\begin{tabular}{r|cc}
    & ${\overline  T}$  & $\ov{u}_3$  \\ \hline
   ${\overline  T^c}$  & $\mu_H$  & $y_2 \bar f_3$.
\end{tabular}
\end{center}
One can see that for $y_2 \bar f_3 \gg \mu_H$, the massless
combination is mostly $ {\ov T}$. For example, if we take 
$\mu_H = 0.3$~TeV, $y_2=1.5$, and $\bar f_3 = 0.6$~TeV, then
\beq
\label{eq:righttop}
\bar t_R = 0.95\, {\ov T}- 0.32\, \ov{u}_3.
\eeq

As the left-handed and right-handed top quarks lie mostly in
$\lambda$ and $ {\ov T}$ respectively, the top Yukawa coupling
predominantly comes from the gaugino interaction,
\beq 
\hat g_5 H_1^\dagger \lambda {\ov T}~. 
\eeq 
This explains why
the top Yukawa coupling is ${\cal O}(1)$. Note that the top quark gets its mass mostly from
$H_1$, which is the same Higgs field that gives masses to the {\it
down\/} type quarks and the charged
leptons, while the up and charm quarks get masses from $H_2$. This
is because unlike the holomorphic superpotential, the
gaugino interaction involves the conjugate of the Higgs field. The
top quark also receives a small contribution to its mass from the
$y_2$ term and nonrenormalizable terms, which can account for the
small mixings with light quarks.

It is also interesting to note that if $\Phi_j$, $\overline{\Phi}_j$ fields have additional interactions, for instance, in the case that $\mu_2$, $\mu_3$ may come from the VEV of some other field, the Higgs quartic coupling can receive a large tree-level correction, 
proportional to $\hat g_5^2$, 
compared with the MSSM value after integrating out the heavy states.

Since we unify the right-handed top quark with the Higgs field in the
same  $SU(5)$ multiplet, one may ask whether $R$-parity is conserved in this model.
We have already seen that the scalar pieces of $\Phi_2$ and
${\ov \Phi_2}$ that have VEVs must have
even $R$-parity.  We have implicitly assumed above that
the color triplet fermions in $\Phi_j$,
${\ov \Phi}_j$, and ${\ov H}$ have even $R$-parity and
therefore that the scalar partners have odd $R$-parity.
It is easy to check that a modified $R$-parity with a
twist $P= (-1,-1,-1,1,1)$ in the $SU(5)$ sector will
satisfy the above constraints.  This twisted $R$-parity
plays the same role as the usual $R$-parity in MSSM and it is conserved in
all the interactions.  More explicitly the bosonic and
fermionic components of ${\ov H}$ are
\beq
{\ov H} \supset ({\tilde {\ov T}}, H_1)+ \theta ( \ov{T}, {\tilde H}_1)~.
\eeq
Similarly if we use a subscript $s$ and $a$ to
denote the $SU(3)_C$ singlet and adjoint pieces of $\Phi_3$,
and  $s$ and $a$ for the $SU(2)_L$  singlet and adjoint pieces of $\Phi_2$ we have
\beq
\Phi_{j} \supset (\Phi_{js},\Phi_{ja},{\tilde \Phi}_{jt}) +\theta
({\tilde \Phi}_{js},{\tilde \Phi}_{ja}, \Phi_{jt})~. 
\eeq
Similarly, there is a new baryon number symmetry which is a linear combination of the original baryon number (with $\Phi_3$, $\overline{\Phi}_3$ carrying original baryon number $-1/3$, $+1/3$ respectively) and a gauge transformation $2(V+aT_{24})$. It is left unbroken by the VEVs, so there is no dangerous proton decay or neutron-anti-neutron oscillation induced.


The couplings of $W'$, $B^{\prime}$, and $B^{\prime\prime}$ to the light SM fermions are suppressed by $\hat{g}_2/\hat{g}_5$, $\hat g_{1H}/\sqrt{15}\hat g_5$ and $\hat g_{1H}/\hat g_{1V}$ relative to the SM $W$ and $B$ gauge bosons respectively. The experimental constraints on $Z'$ put a lower bound on $m_{W'}$ at about 800 GeV. The more stringent constraints come from the mixing of $W$, $B$ and $W'$, $B'$, $B''$ due to the Higgs VEVs because the two Higgs doublets and the light fermions transform under different gauge groups. In terms of the usual $S$, $T$ parameters~\cite{peskin}, the strongest constraint comes from $T$ for $\hat g_{1V}\gtrsim 2-3$ and it depends only on $f_2^2 +\bar f_2^2$. For a light Higgs ($\sim 115$~GeV) $f_2^2 +\bar f_2^2$ is required to be $\gtrsim (3~{\rm TeV})^2$. For a heavier Higgs the bound is weaker and $f_2^2 +\bar f_2^2$ might be as low as $\sim (2~{\rm TeV})^2$. Another constraint comes from the corrections to the $Z \to b \bar b$ vertex. The $b_L$ quark in this model is a linear combination of several fields and is mostly the gaugino of $SU(5)$. It couples to the heavy gauge bosons differently from $d_L,\, s_L$. It induces a correction to $Z b_L \bar b_L$ coupling after Higgs VEVs mix $Z$ with $W'_3$, $B^{\prime}$, and $B^{\prime\prime}$. Requiring that $\delta g_{Lb}/g_{Lb} < 0.6\%$ puts a lower bound of $\sim$1.6 TeV for $m_{W'}$. Note, however, that these are indirect constraints. It is possible that these contributions are cancelled by some additional contributions. It is still quite possible that $\vec{Q}$ has a mass below 2 TeV. 

The model predicts the existence of new gauge bosons $G',\, W',\,
B^{\prime}, B^{\prime\prime}$ and several new fermions $t',\, b'$ around the same scale as the spin-1
top superpartner $\vec{Q}$. The masses of the gauge bosons are closely related through Eqs.~(\ref{eq:gprime})--(\ref{eq:vtopmass}).
For heavy $t'$ and $b'$ quarks, there are three doublets and one singlet. Their masses depend on the model parameters. For the set of the parameters considered earlier above Eq.~(\ref{eq:lefttop}), (\ref{eq:righttop}) and $\hat g_3 =2.0$, $\hat g_2 =0.75$, $\hat g_{1H} =0.36$, $\hat g_{1V} =3.5$ at $\sim$ 2 TeV (to produce the measured values of SM gauge couplings), the sample spectrum for the heavy gauge bosons and fermions is listed below.
\beq
\begin{tabular}{c|ccccccccc}
 & $G'$ & $  W'$  & $
B^\prime$ &$B^{\prime\prime}$& ${\vec Q}$ & $ Q' $ & $Q''$ &$Q'''$ & $\ov{T}'$
\\ \hline $M/ \mbox{TeV}$ &  $1.7$  & $3.2$ & $0.83$ & $2.6$& $2.0$ &
$0.65$ & $3.0$ & $5.8$ & $0.95$ 
\end{tabular}
\eeq

The spectrum of superpartners depends on the soft-SUSY-breaking terms. The superpartners of the light fermions can have masses in the multi-TeV range without affecting the naturalness of electroweak symmetry breaking because of the small Yukawa couplings. The SUSY-breaking masses of $\Phi_{2,3},\, \ov{\Phi}_{2,3}$ should also be in the TeV range in order to induce symmetry breaking VEVs of the TeV scale. We assume that all soft-SUSY-breaking scalar masses except those of $H$ and $\ov{H}$ are large ($\gtrsim$ a few TeV). In this case, most of the scalar superpartners are beyond the reach of the LHC. On the other hand, the soft masses of $H$ and $\ov{H}$ are relevant for stabilizing the electroweak scale as they contain the Higgses and the right-handed top quark. They need to be at $\sim$ 1 TeV or below. Electroweak gauginos also give significant contributions to the Higgs masses. We assume that the SUSY-breaking gaugino masses are of the order a few hundred GeV, smaller than (most of) the soft scalar masses. This can be naturally achieved if there is an approximate $R$ symmetry in the SUSY breaking sector. 

\begin{figure}[htb]
\begin{center}
\includegraphics[width=0.9\hsize]{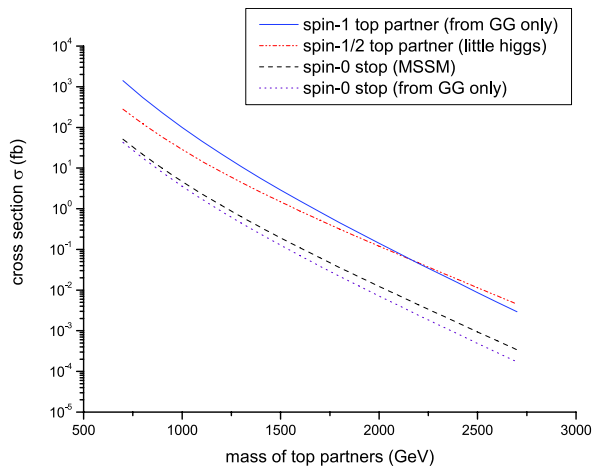}
\end{center}
\caption{Total cross-sections vs. mass of top partners\label{fig:crosssection}}
\end{figure}

With the spectrum assumed above, the superpartners of the SM particles that can be seen at the LHC are the spin-1 partner of the left-handed top-bottom doublet, the scalar partner of the right-handed top, gauginos of the SM gauge group, and Higgsinos. In addition, we expect to see new gauge bosons, $t'$, $b'$, and some of their superpartners. We will focus on the spin-1 partner of the top and bottom doublet. At the LHC, the main production mechanism is $GG \to \vec{Q} \vec{Q}^*$. The processes with $q \bar q$ initial states are suppressed by destructive interference between $G$ and $G'$ exchanges because $q,\bar q$ and $\vec{Q}$ transform under different UV gauge groups. The dominant production cross-sections of the spin-1 and spin-0 top partners of this model from gluons are shown in Fig.~\ref{fig:crosssection} as functions of their masses. In comparison, the cross-sections of the (right-handed) top squark in MSSM and fermionic top partner in little Higgs models are also shown.

The total cross-sections of the top partners have been used as a way to distinguish little Higgs models from SUSY models as the total cross-section of the fermionic top partner is much larger than that of a scalar partner for the same mass~\cite{toppartners}. However, we see that the spin-1 top partner can have an even larger total cross-section for the same mass.

\begin{figure}[htb]
\begin{center}
\begin{tabular}{lcr}
\includegraphics[width=0.35\hsize]{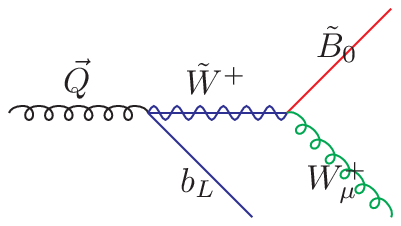}
\end{tabular}
\end{center}
\caption{A decay chain for the spin-1 top partner.\label{fig:decay}  }
\end{figure}

With a similar spectrum the $\vec{Q} $ decay is similar to the usual top squark decay . If the gauginos are light, it will decay to a gaugino and a top or bottom quark, and eventually end up with a lightest $R$-parity odd particle through cascade decays. One such possible decay channel is shown in Fig.~\ref{fig:decay}. It may have a different angular distribution for its decay products compared with the usual top squark, but it would be very challenging to tell at the LHC.

In conclusion, we have shown that a spin-1 top partner requires an extended gauge symmetry 
which can explain why one quark is much heavier than
the others, and provide a larger Higgs quartic coupling than in the MSSM. The spin-1 top partner
thus provides a interesting (and experimentally challenging) new scenario for LHC physics.

\acknowledgments
We thank K.~Agashe and C.~Csaki for pointing out an extra $U(1)$ symmetry in this model.
This work was supported in part by U.S. DOE grant No. DE-FG02-91ER40674.

\end{document}